\documentclass[apj]{emulateapj}

\newcommand{\degree}{\hbox{$^\circ$$$}}

\shorttitle{Stellar populations and structure of Leo T}
\shortauthors{de Jong et al.}

\begin{document}

\title{The structural properties and star formation history of Leo T
  from deep LBT photometry\altaffilmark{7}}

\author{J. T. A. de Jong\altaffilmark{1}, J. Harris\altaffilmark{2},
  M. G. Coleman\altaffilmark{1}, N. F. Martin\altaffilmark{1},
  E. F. Bell\altaffilmark{1}, H-W. Rix\altaffilmark{1},
  J. M. Hill\altaffilmark{3}, E. D. Skillman\altaffilmark{4},
  D. J. Sand\altaffilmark{2}, 
  E. W. Olszewski\altaffilmark{2}, D. Zaritsky\altaffilmark{2},
  D. Thompson\altaffilmark{3}, E. Giallongo\altaffilmark{5},
  R. Ragazzoni\altaffilmark{6}, A. DiPaola\altaffilmark{5},
  J. Farinato\altaffilmark{6}, V. Testa\altaffilmark{5},
  J. Bechtold\altaffilmark{2} }

\email{dejong@mpia.de}

\altaffiltext{1}{Max-Planck-Institut f\"{u}r Astronomie,
  K\"{o}nigstuhl 17, D-69117 Heidelberg, Germany}
\altaffiltext{2}{Steward Observatory, University of Arizona, 933 North
  Cherry Ave., Tucson AZ 85721-0065, United States}
\altaffiltext{3}{Large Binocular Telescope Observatory, University of
  Arizona, 933 North Cherry Ave., Tucson, AZ 85721, United States}
\altaffiltext{4}{Astronomy Department, University of Minnesota,
  Minneapolis, MN 55455, United States}
\altaffiltext{5}{INAF, Osservatorio Astronomico di Roma, via di
  Frascati 33, I-00040 Monteporzio, Italy}
\altaffiltext{6}{INAF, Osservatorio Astronomico di Padova, vicolo
  dell'Osservatorio 5, I-35122 Padova, Italy}
\altaffiltext{7}{Based on data acquired using the
  Large Binocular Telescope (LBT). The LBT is an international
  collaboration among institutions in the United States, Italy and
  Germany. LBT Corporation partners are: The University of Arizona on
  behalf of the Arizona university system; Instituto Nazionale di
  Astrofisica, Italy; LBT Beteiligungsgesellschaft, Germany,
  representing the Max-Planck Society, the Astrophysical Institute
  Potsdam, and Heidelberg University; The Ohio State University, and
  The Research Corporation, on behalf of The University of Notre Dame,
  University of Minnesota and University of Virginia.}

\begin{abstract}
We present deep, wide-field $g$ and $r$ photometry of the transition
type dwarf galaxy Leo T, obtained with the blue arm of the Large
Binocular Telescope. The data confirm the presence of both very young
($<$1 Gyr) as well as much older ($>$5 Gyr) stars. We study the
structural properties of the old and young stellar populations
by preferentially selecting either population based on their color and
magnitude. The young population is significantly more
concentrated than the old population, with half-light radii of
104$\pm$8 and 148$\pm$16 pc respectively, and their centers are
slightly offset. Approximately 10\% of the total stellar mass is
estimated to be represented by the young stellar
population. Comparison of the color-magnitude diagram (CMD) with
theoretical isochrones as well as numerical CMD-fitting suggest that
star formation began over 10 Gyr ago and continued in recent
times until at least a few hundred Myr ago. The CMD-fitting results
are indicative of two distinct star formation bursts, with a quiescent
period around 3 Gyr ago, albeit at low significance. The results are
consistent with no metallicity evolution and [Fe/H]$\sim -$1.5 over
the entire age of the system. Finally, the data show little if
any sign of tidal distortion of Leo T.
\end{abstract}


\section{Introduction}
\label{sec:intro}

Leo T clearly stands out from the large number of new dwarf galaxies recently
discovered \citep{UMaI,Wil1,CVnI,UMaII,Boo,5pack,LeoT,BooII} around the Milky
Way using Sloan Digital Sky Survey \citep[SDSS,][]{york00,dr5} data. It is by
far the most distant, located approximately 420 kpc from the Galaxy and
probably not bound to it. Furthermore, it is the only one that contains very
young ($<$1 Gyr) stars, apart from an old or intermediate age population, and
has detected HI associated with it \citep{LeoT}. In terms of its properties it
seems to be a transitional system, as it combines the round, regular structure
of dwarf spheroidal galaxies with the presence of gas and recent star
formation common in dwarf irregulars.

The other known so-called transition type dwarf galaxies, DDO 210, Phoenix,
Pisces, Antlia, Pegasus and Leo A \citep[e.g.][]{mateo98,grebel01}, are also
isolated Local Group members, but much brighter than Leo T. Because of
its large distance from the Milky Way, it is unlikely that Leo T has
been significantly affected by the tidal forces of the Milky Way,
suggesting that its current low luminosity is intrinsic. From the
velocity dispersion of 19 red giants \cite{simongeha} infer a dark
halo mass of $\sim 10^7 M_\sun$ and a mass-to-light ratio of
$\sim 140~ M_\sun/L_{V,\sun}$ for Leo T. From HI observations,
\cite{ryanweber08} infer a lower limit for the total dynamical mass of
$\sim 3\times10^6 M_\sun$ and a mass-to-light ratio of $\gtrsim 56~
M_\sun/L_{V,\sun}$. They also detect two distinct components in the
neutral hydrogen, one cold ($\sim$500 K) and one warm ($\sim$6000 K).
It is interesting to see that such a low luminosity system still
contains gas and very young stars at the current epoch.

In this work we present deep photometry of Leo T, obtained with the
Large Binocular Telescope. We use these data, the deepest data available
on Leo T, to study its stellar populations and structural properties.
The outline of this paper is as follows. In
\S \ref{sec:data} we will first describe the data used in this
work. Based on color-magnitude diagram (CMD) morphology we describe the
stellar populations present in Leo T in \S \ref{sec:pops}. The
structural properties of Leo T are analyzed in \S
\ref{sec:struct} and in \S \ref{sec:sfh} we use CMD-fitting
techniques to constrain its star formation history and metallicity
evolution. Finally, our conclusions are presented in \S
\ref{sec:conclusions}.

\section{Data}
\label{sec:data}

\begin{figure*}[t]
\epsscale{0.8}
\plotone{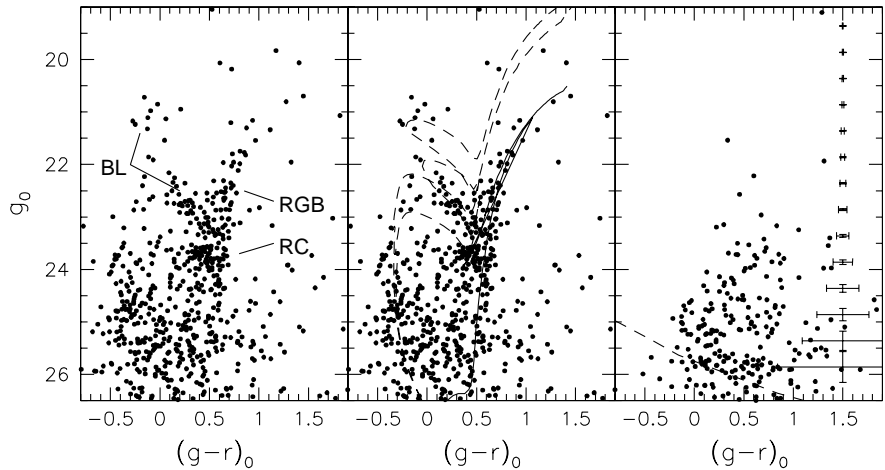}
\caption{ {\it Left:} CMD of stars within 1.4\arcmin~ of the center of
Leo T, with the features discussed in \S \ref{sec:pops}
indicated. {\it Center:} Same CMD as in the left panel, now
with isochrones overlayed to outline the two distinct stellar
populations. The dashed lines are for 400 and 650 Myr from left
to right, the solid line for 10 Gyr and all assume [Fe/H]$=-$1.7 and
m-M=23.1 mag. {\it Right:} CMD of an annulus of radius 6\arcmin~
centered on Leo T of equal area as the target region, which
serves as a control field. Photometric errors are indicated on the
right and the dashed line shows the 50\% completeness limit.  }
\label{fig:cmds}
\end{figure*}

\begin{figure*}[ht]
\epsscale{1.0}
\plotone{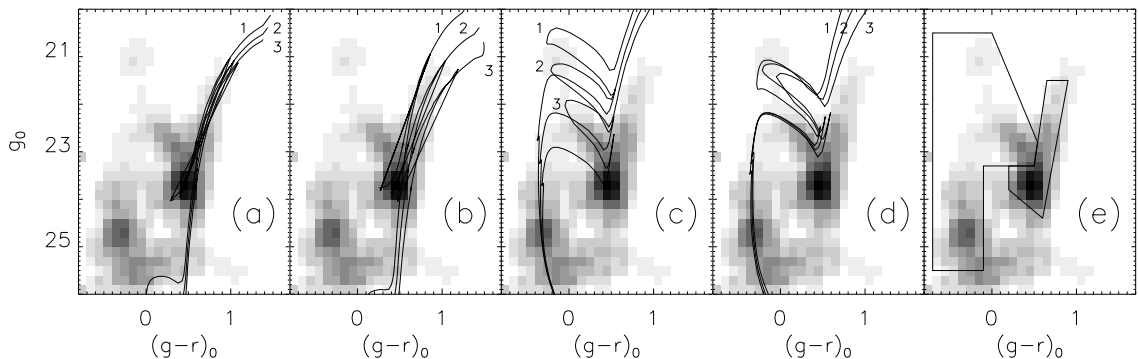}
\caption{Background-subtracted Hess diagrams of Leo T with different
  overlays: 
{\it (a):} isochrones of 5 (1), 8 (2) and 12.5 Gyr (3) with [Fe/H]$=-$1.7 dex
{\it (b):} isochrones of 8 Gyr for metallicities of [Fe/H]$=-$2.3 (1),
  $-$1.7 (2) and $-$1.3 dex (3)
{\it (c):} isochrones of 250 (1), 400 (2) and 650 (3) Myr with [Fe/H]$=-$1.7 dex
{\it (d):} isochrones of 400 Myr for metallicities of [Fe/H]$=-$2.3 (1),
  $-$1.7 (2) and $-$1.3 dex (3)
{\it (e):} CMD selection boxes to select preferentially either stars
  belonging to the young (left-hand box) or the old (right-hand box)
  stellar population.
The Hess diagrams were created by subtracting the scaled Hess diagram
  of the region outside 6\arcmin~ from the center of Leo T from that
  of the region within 1\farcs4, after which a boxcar smoothing of
  width 2 is applied.
}
\label{fig:hess}
\end{figure*}

Deep $g$- and $r$-band photometry was obtained with the blue channel
of the Large Binocular Camera \citep[LBC,][]{ragazzoni06,giallongo07}
mounted at the prime focus of the Large Binocular Telescope
(LBT). Recently commissioned, the LBT consists of two 8.4 meter
mirrors on a single mount \citep{hill06} and is located on Mount
Graham in Arizona, USA.  The LBC is a wide-field imager consisting of
four 2048$\times$4608 pixel chips, providing a field-of-view of
23\arcmin$\times$23\arcmin~ with a pixel size of 0\farcs23. The data,
obtained as part of the LBT Science Demonstration Time program, were
taken on 2007 March 21 under seeing conditions of
1\farcs0$-$1\farcs1. In each filter, four exposures of five minutes
each were taken with small offsets to help with bad pixel and cosmic
ray rejection. One $r$ band frame was not used for the current work
because of poor guiding.

Data reduction was performed using the reduction pipe-line developed
by \cite{coleman07}, based on standard routines in the IRAF package
{\it mscred}. The images were trimmed, bias-subtracted and
flat-fielded using the combined twilight flats. Because of the `fast'
focal ratio ($f$/1.14) of the LBT primary mirrors there is significant
field distortion (1.75\%) near the edges of the field-of-view. This
distortion was removed with a quadratic radial correction to an
accuracy of $\sim$0\farcs2. Because of the effectively different
pixel scale, the photometry must also be flattened before co-adding
the single chip images into mosaics. After that the images were
astrometrically registered (accuracy $\sim$0\farcs1 w.r.t. the SDSS
catalog) and median combined to produce the final science frames.

Detection of stars and photometry was performed using the PSF-fitting
photometry package DAOPHOT. Because the observations were done in filters that
are very close to the SDSS $g$ and $r$ filters, the photometry was calibrated
to SDSS without the need for using standard star observations or atmospheric
extinction corrections. The accuracy of the zeropoint calibration was $\delta
g \sim$0.03 mag and $\delta r \sim$0.02 mag. Based on the dust extinction maps
from \cite{sfd} the photometry was corrected star-by-star for foreground
extinction; the extinction in the Leo T field varies between E(B-V)$=$0.027 and
0.041 mag, with a mean of E(B-V)$=$0.035 mag. For clarity, we will refer to the
extinction-corrected magnitudes as $g_0$ and $r_0$ in the remainder of this
paper. The photometric uncertainties and completeness were determined with
artificial star tests. 1600 artificial stars were placed at 0.25 magnitude
intervals between 16th and 29th magnitude in the $g$ and $r$ images and
subsequently detected and photometered using DAOPHOT. The photometric
uncertainty is taken to be the dispersion of the recovered magnitudes around
the mean magnitude.

The resulting color-magnitude diagram (CMD) of the central 1.4\arcmin,
the half-light radius according to \cite{LeoT},
of Leo T is shown in panels (a) and (b) of Figure \ref{fig:cmds}. In panel (c)
the CMD of an annulus of equal area at a radius of 6\arcmin~ from the
center of Leo T is shown; the photometric errors and 50\% completeness
line are also indicated in this panel.

\section{Stellar populations}
\label{sec:pops}

Before turning to an analysis of the structure of the Leo T dwarf
galaxy, we spend some time describing its two main stellar populations
already seen by \cite{LeoT}. The structural properties of these two
populations will then be studied separately. 

In the CMD in panel (a) of Figure \ref{fig:cmds} several features are
readily visible:
\begin{itemize}
\item a red giant branch (RGB) starting at
($g$-$r$,$g$)$_0$$\sim$(1.3,21) and going all the way down to
  $\sim$(0.5,26)
\item a red clump (RC) feature around ($g$-$r$,$g$)$_0$ $\sim$(0.4,23.7)
\item a ridge of stars between ($g$-$r$,$g$)$_0$ $\sim$(0.5,23.3) and
(0.0,22.5), reminiscent of a blue loop (BL)
\item a small group of stars around (0.0,21.0), possibly also related
  to the BL
\item an excess of stars bluer than ($g$-$r$)$_0=$0.0, presumably
  young main-sequence (MS) stars
\end{itemize}

The RGB, RC and BL were also identified by \cite{LeoT}, who attributed
the first two to an intermediate age to old ($\sim$5-12 Gyr) and the
latter to a very young ($<$1 Gyr) population. In Figure \ref{fig:hess}
we present smoothed, background-subtracted Hess diagrams \citep[2-D
histograms of stellar density in the color-magnitude plane,][]{hess24}
of our photometry in the central region of Leo T. In \S
\ref{sec:sfh} we will redetermine the distance to Leo T, but for the
moment we assume the distance modulus of 23.1 mag from the discovery
paper (based on HB luminosity and CMD-fitting) and overlay three
isochrones from \cite{girardi04} for ages of 5, 8, and 12.5 Gyr and a
metallicity of [Fe/H]$=-$1.7 dex in panel (a). It is clear that all three
isochrones fit the RGB and RC very well and that photometry down to
the turn-off ($g\sim$27 for 12.5 Gyr) will be necessary to constrain
the age of this older stellar population. In panel (b) three 8 Gyr
isochrones of [Fe/H]$=-$2.3, $-$1.7, and $-$1.3 dex are overlayed on
the Hess diagram, indicating that both the average color of the RC and
the slope of the RGB favor a metallicity of $-$1.7 dex.

Isochrones for much younger ages of 250, 400, and 650 Myr and [Fe/H]=-1.7 dex
are overplotted on the Hess diagram in panel (c) of Figure \ref{fig:hess}.
This clearly implies that the stars bluewards of ($g$-$r$)$_0$=0.0 are young
MS stars, and that the BL feature and group of stars around (0.0,21.0) are
indeed helium-burning BL stars belonging to the same population. The 250 Myr
isochrone seems to set a lower limit on the age, as its turn-off and BL seem
too bright for the data. The 650 Myr isochrone shows that young stars also
contribute to the RC, increasing the uncertainty in distance and age
measurements using the RC luminosity, as these are age and metallicity
dependent \citep{girardisalaris}.  That the metallicity of the young stars is
very difficult to determine is illustrated in panel (d), where 400 Myr
isochrones with [Fe/H] of $-$2.3, $-$1.7, and $-$1.3 are overplotted. The
regions of the CMD that are sensitive to the metallicity differences are too
sparsely populated to make this measurement.

\begin{figure}
\epsscale{1.0}
\plotone{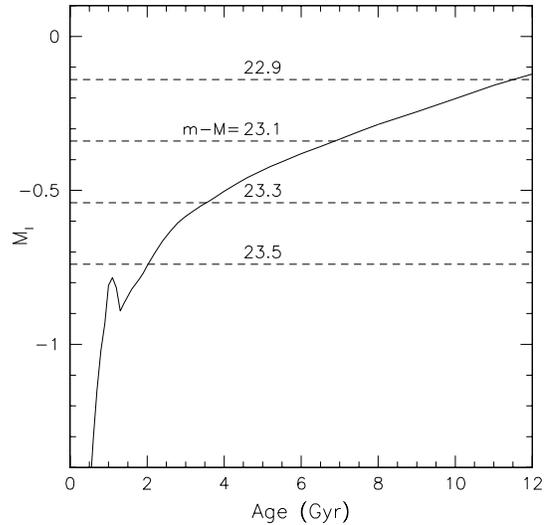}
\caption{
Relation between age and distance modulus determined from RC
luminosity. The solid line gives the absolute $I$-band magnitude of
RC stars as a function of age for a metallicity of [Fe/H]$=-$1.7,
following \cite{girardisalaris}. Dashed lines indicate the absolute
magnitude of the RC in Leo T assuming distance moduli of 22.9, 23.1,
23.3 and 23.5, from top to bottom. The intersection of each dashed
line with the solid line gives the inferred age for each distance
modulus.
}
\label{fig:redclump}
\end{figure}

While the luminosity and color of the RC can be used to constrain population
parameters such as distance, age and metallicity, there are degeneracies and
solutions are generally not unique \citep{girardisalaris}. The presence of a
RC, rather than a more extended HB, in itself implies that the stars are not
both old and metal-poor ([Fe/H]$<-$2). However, since the metallicity of the
older stars is relatively well-constrained by the RGB slope, we can use the RC
luminosity to determine either the distance or the age of the older stars,
using the theoretical RC absolute $I$-band luminosities from
\cite{girardisalaris}. From our data we determine the location of the RC by
iteratively finding the mean color and magnitude of stars in a 0.25$\times$0.5
mag box centered on the RC. Using different starting color and magnitude
values around the approximate center we find $g_{0,RC}$=23.71$\pm$0.03 and
$(g-r)_{0,RC}$=0.43$\pm$0.02, where the errors are determined from bootstrap
resampling tests.  Using the photometric equations determined by Lupton
(2005)\footnote{http://www.sdss.org/dr5/algorithms/sdssUBVRITransform.html}
this translates to $m_I$=22.76$\pm$0.04 mag. The interplay between age and
distance for the brightness of the RC, at a metallicity of [Fe/H]$=-1.7$, is
illustrated in Figure \ref{fig:redclump}, where we plot the absolute magnitude
of RC stars as a function of age \citep{girardisalaris}.  Assuming a distance
modulus of 23.1, the RC luminosity implies an average age of the older stars in
Leo T of $\sim$7 Gyr. Alternatively, considering that the age of the stars
might vary from 5 to 13 Gyr, as indicated by the left-most panel of Figure
\ref{fig:hess}, this implies that the distance modulus lies between 22.9
and 23.2 magnitudes.

\section{Structural properties}
\label{sec:struct}

\begin{figure}
\epsscale{1.0}
\plotone{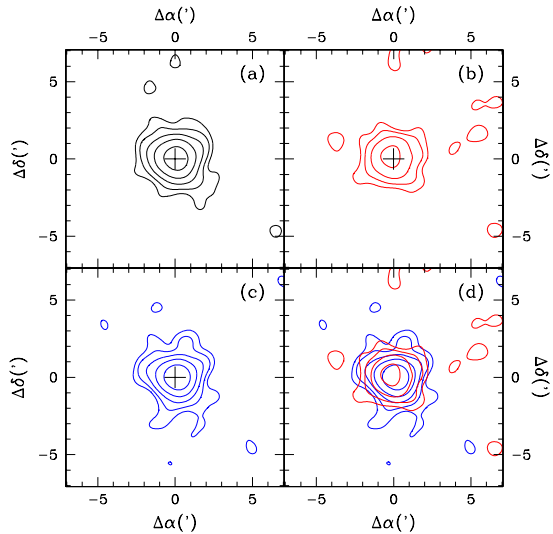}
\caption{
Stellar density distributions in Leo T for: 
{\it (a):} all stars;  
{\it (b):} the old stars;  
{\it (c):} the young stars;  
{\it (d):} both old (red contours) and young (blue contours) stars.
These density maps are smoothed with a 0\farcm6 Gaussian smoothing
kernel and the contours correspond to densities of 1.5$\sigma$,
3$\sigma$, 6$\sigma$, 12$\sigma$, and 25$\sigma$ above the
background. As a reference, the cross in panels (a) through (c)
indicates the center of the distribution of all stars.
}
\label{fig:densities}
\end{figure}

\begin{figure}
\epsscale{1.0}
\plotone{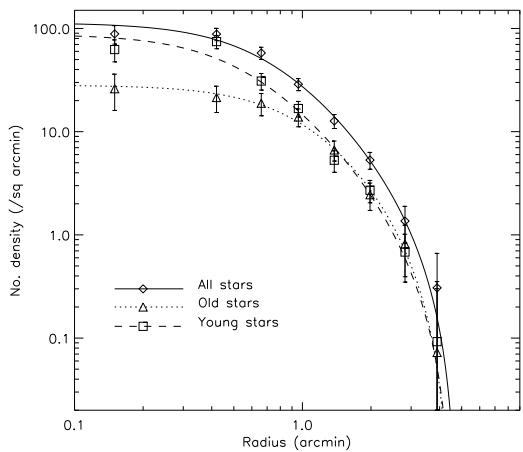}
\caption{ 
Background-corrected radial profiles of Leo T, determined
from the CMD-selected young (squares), old (triangles), and all
(diamonds) stars. The lines represent the best-fitting King profiles,
determined for each subset of stars separately: the dashed line for
the young, the dotted for the old and the solid for all stars.
}
\label{fig:kingfits}
\end{figure}

The stellar sources within 1.4\arcmin~ from the center of Leo T are
divided in two subsets corresponding to the young ($<$1 Gyr)
and the old ($>$5 Gyr) stellar populations described in \S
\ref{sec:pops}. This is done using the CMD selection boxes outlined in panel
(e) of Figure \ref{fig:hess}; by combining these two subsets we create
a third subset of both old and young Leo T candidate members.
In this section we study the structural properties of these three
subsets, which we will refer to as `all', `young', and `old' stars,
and which contain 2247, 854, and 1393 stars, respectively.

The central coordinates of Leo T were determined for the three subsets by
iteratively calculating the mean center of the distribution of stars
on the sky within an aperture of 6\arcmin~ centered on the assumed
center, until the offset between subsequent iterations was 
zero.  In Table \ref{tab:structprop} the RA and Dec for each set of
stars is listed. Errors in the obtained centers were determined using
bootstrap resampling.  There is a small offset of 0\farcm3, or 35
pc, between the centers of the distribution of young and old stars,
which is significant at the 2$\sigma$ level.

Figure \ref{fig:densities} displays the surface density of stars as contour
diagrams; panels (a) through (c) are for all stars, the old stars, and the
young stars, respectively, and panel (d) shows the contours of both the old
and young stars. The center of the distribution of all stars is indicated with
a cross in panels (a) through (c), to help identify the apparent offset
between the different subsets of stars. The contours correspond to densities
of 1.5, 3, 6, 12, and 25 $\sigma$ above the background, where $\sigma$ is the
Poisson uncertainty in the background stellar density. Leo T appears to be
very compact and round and there is little sign of any distortion or
extra-tidal features. Fitting of a series of elliptical contours using the
IRAF routine {\it ellipse} shows that for all subsets of stars the ellipticity
is below 0.1 out to a radius of 3\arcmin~ from the center, where the density
of stars drops below 1.5$\sigma$ above the background.

To measure the half-light radii, core radii, and limiting radii of Leo T and
its sub-populations, King profiles and exponential profiles were fit to the
stellar distributions. For each subset of stars, the profiles were fit to the
stellar density in circular annuli of increasing radius and width, centered on
the center found for that subset, as listed in Table \ref{tab:structprop}.
Crowding is not an issue for star counting down to the magnitude limit we use
(25.5 mag, see Figure \ref{fig:hess}).  The results for the exponential
($r_e$), half-light ($r_h$), core ($r_c$), and limiting ($r_t$) radii are
listed in Table \ref{tab:structprop} and the density profiles with the
corresponding best King profile fits are presented in Figure
\ref{fig:kingfits}. The uncertainties on the fit parameters were determined
using bootstrap resampling.  Contrary to the tentative conclusion by
\cite{LeoT} that the young stars seem less concentrated, comparing the density
profiles from our deeper data shows unambiguously ($\sim 3 \sigma$) that the
young stars are more centrally concentrated; their central density is three
times higher than that of the old stars, while the limiting radii are
indistinguishable. This leads to a difference of a factor two in the
concentration parameter $c=r_t/r_c$ between these populations.

\begin{deluxetable*}{lccc}
\tablecaption{Leo T structural properties}
\tablewidth{0pt} 
\tablehead{ \colhead{Parameter} & \colhead{All} & \colhead{Old} & \colhead{Young} }
\startdata
RA & 143\degree.723$\pm$0\degree.001 & 143\degree.727$\pm$0\degree.003 & 143\degree.722$\pm$0\degree.001 \\
RA & 9$^h$34$^m$53.5$^s$$\pm$0.2$^s$ & 9$^h$34$^m$54.5$^s$$\pm$0.7$^s$ & 9$^h$34$^m$53.3$^s$$\pm$0.2$^s$ \\
Dec & 17\degree.051$\pm$0\degree.001 & 17\degree.050$\pm$0\degree.003 & 17\degree.050$\pm$0\degree.001 \\
Dec & 17\degree03\arcmin04\arcsec $\pm$4\arcsec & 17\degree03\arcmin00\arcsec$\pm$12\arcsec & 17\degree03\arcmin00\arcsec$\pm$4\arcsec \\
$r_e$ (\arcmin) & 0.59$\pm$0.04 & 0.73$\pm$0.08 & 0.51$\pm$0.04 \\
$r_h$ (\arcmin) & 0.99$\pm$0.06 & 1.22$\pm$0.13 & 0.86$\pm$0.07 \\
$r_c$ (\arcmin) & 0.68$\pm$0.08 & 1.05$\pm$0.27 & 0.54$\pm$0.08 \\
$r_t$ (\arcmin) & 4.8$\pm$1.0   & 4.5$\pm$1.1   & 4.6$\pm$1.1   \\
$r_e$ (pc)      & 72$\pm$5      & 89$\pm$10     & 62$\pm$5      \\
$r_h$ (pc)      & 120$\pm$7     & 148$\pm$16    & 104$\pm$8     \\
$r_c$ (pc)      & 82$\pm$10     & 127$\pm$33    & 65$\pm$10     \\
$r_t$ (pc)      & 580$\pm$120   & 550$\pm$130   & 560$\pm$130   \\
$c$             & 7.1$\pm$1.7   & 4.3$\pm$1.5   & 8.6$\pm$2.4   \\
$M_V$ (mag)     & -8.0          & -7.5          & -6.9          \\
m-M$_{RC}$ (mag) & 23.1$\pm$0.2 & ~ & ~ \\
m-M$_{MATCH}$ (mag) & 23.0$\pm$0.2 & ~ & ~ \\
\enddata
\label{tab:structprop}
\end{deluxetable*}

For calculating the total $V$-band luminosity of Leo T we convert the $g_0$
and $r_0$ magnitudes of all stars to $B$ and $V$ using again the photometric
transformations from Lupton (2005). The total luminosity in stars within a
5\arcmin~ aperture centered on Leo T (five times the half-light radius
determined above), integrated between $V$-band magnitudes of 20.5 and 24.5
($\sim$1 magnitude below the RC) is $M_V=-$7.1 mag. This includes both old and
young stars, but is not corrected for stars fainter than $V=$24.5 and should
therefore be taken as a minimum luminosity for Leo T. In this calculation a
correction for the field stars has been done using the area outside a
6\arcmin~ radius. Since the MSTO of the young stars is at roughly V$=$24.5,
most of the luminosity of the young population should be included in this
estimate. For the older stars, we use the 10 Gyr, [Fe/H]$=-$1.5 luminosity
function (LF) from \cite{dotter07}, assuming a Salpeter initial mass function
\citep{salpeter} and find that the magnitude interval we use should contain
approximately 38\% of the total light; for the young stars we estimate this
should be $\sim$60\% of the total light. Using the CMD box to select stars
belonging to the older population, we get $M_{V,old}=-$6.4 and
$M_{V,young}=-$6.3, roughly equal values; including the LF correction yields
$M_{V,old}=-$7.5 mag and $M_{V,young}=-$6.9 mag. Combining all this, we
estimate the total luminosity of Leo T to be $M_{V,total} \simeq -$8.0 mag.
These fractions of the contributed light translate to a stellar mass fraction
of the young stars of $\sim$10\%, with some error, especially due to
uncertainties in the initial mass function.

\section{Star formation history}
\label{sec:sfh}

CMD-fitting techniques can be used to constrain the stellar population
properties of dwarf galaxies in detail \citep[e.g.][]{gallart96,
tolstoy96, aparicio97, dolphin97, holtzman99, olsen99, hernandez00,
harris01}. Here we apply two different CMD-fitting packages, StarFISH
\citep{harris01} and MATCH \citep{match}, to our photometry of stars
within 1\farcm4 of the center of Leo T and constrain its star
formation history and metallicity evolution.

Both CMD-fitting packages are based on the same principles, although the exact
implementation differs slightly. Based on theoretical isochrones, here from
\cite{girardi04}, artificial CMDs are constructed for different combinations
of stellar population parameters such as distance, age, metallicity, initial
mass function and binary fraction. Convolving these theoretical CMDs with the
photometry errors and completeness determined from the artificial star tests
yields realistic model CMDs that can be compared to the data. Converting the
models and data to Hess diagrams, enables a direct pixel to pixel comparison.
The fit result is the best-fitting linear combination of model CMDs with the
scaling of each model being the SFR for the corresponding stellar population
bin. To correct for the contamination by foreground stars, a control CMD can
be provided, which is then used as an additional model CMD in the fit. Here we
use the stars outside a 6\arcmin~ radius from the center of Leo T for
the control CMD.

For the SFH fits with MATCH we chose a set of metallicity bins,
centered at [Fe/H] of $-$2.2, $-$1.7, $-$1.3 and $-$0.8, with a width of 0.5
dex. Two sets of age bins were adopted, one with bin-widths of $\Delta
\log(t)=0.3$ and one with bin-widths of $\Delta \log(t)=0.4$ dex,
running from $\sim$10 Myr to $\sim$16 Gyr. Stars with colors
$-1.0<(g$-$r)_0<2.0$ and brighter than $g_0=26$ and $r_0=25$ were fit,
with a Hess diagram bin size of 0.15 in magnitude and 0.08 in color.

\begin{figure}
\epsscale{1.0}
\plotone{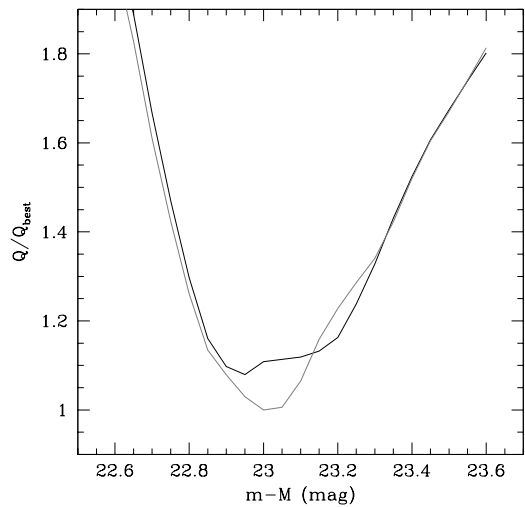}
\caption{Goodness-of-fit relative to that of the best fit for MATCH
  SFH fits assuming different distance moduli. The grey line is for
  age bins of 0.3 log-year width, the black line for 0.4 log-year bins.
}
\label{fig:matchdistances}
\end{figure}

\begin{figure}
\epsscale{1.0}
\plotone{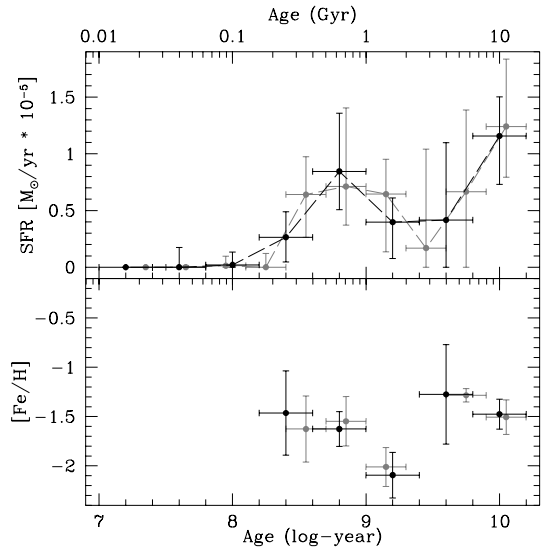}
\caption{ Star formation rate and metallicity evolution of
  Leo T from fits with MATCH. {\it Upper panel: } star
  formation as function of time for 0.3 log-year age bins (grey
  points) and 0.4 log-year bins (black points). Horizontal error bars
  indicate the bin width. {\it Lower panel: } metallicity as function
  of time with gray scale and horizontal error bars as in the upper
  panel. The metallicities are the star formation rate-weighted
  average.  }
\label{fig:matchsfhamr}
\end{figure}

\begin{figure}
\epsscale{1.0}
\plotone{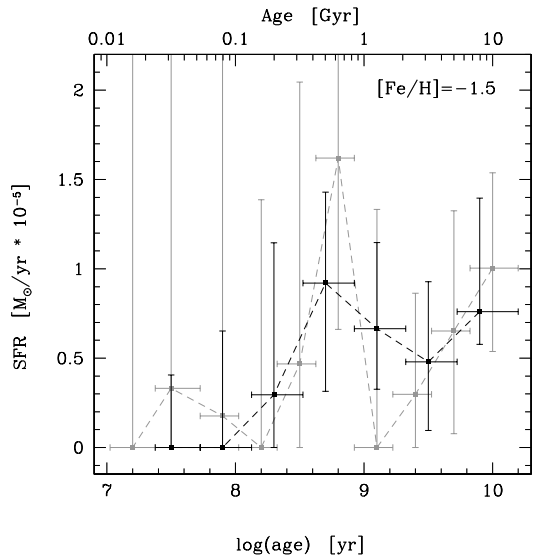}
\caption{The SFH solution from the StarFISH fit.  Here the SFH is
  expressed as the $SFR(t)$ for a metallicity of [Fe/H]=$-$1.5. As in
  Figure~\ref{fig:matchsfhamr}, the solution with $\Delta
  \log(age)$=0.4 is shown in black, and the solution with $\Delta
  \log(age)$=0.3 is shown in grey. }
\label{fig:starfish}
\end{figure}

With foreground extinction corrected based on the dust extinction maps of
\cite{sfd} and age and metallicity fit by the software, the remaining
parameter is the distance to Leo T.  In order to constrain this parameter and
to determine how sensitive the best-fitting SFHs are to the adopted distance,
fits were run for a range of distance moduli between 22.6 and 23.6, at
intervals of 0.05 magnitudes. Figure \ref{fig:matchdistances} shows the fit
qualities for these fits, normalized by that of the best fit. The two
different age binning schemes give a slightly different best distance. The
overall best fit is obtained with the smaller age bins, which is to be
expected since that binning scheme provides more age bins and thus more
degrees of freedom. Taking the expected variance in the fit quality into
account, the statistically ``good'' fits can be determined. These fits have
distance moduli ranging from 22.8 to 23.2, thus giving a distance estimate of
m-M$=$23.0$\pm$0.2. Based on the RC luminosity calculated in section
\ref{sec:pops}, a distance modulus of 23.0 implies an average age for the
stars in Leo T of $\sim$9 Gyr (Figure \ref{fig:redclump}).  Using the fits
with distance moduli between 22.8 and 23.2, the SFH and metallicity evolution
were determined. In Figure \ref{fig:matchsfhamr} the relative star formation
rate and the metallicity are plotted as function of time.

For the StarFISH fit, we matched the input parameters as closely as
possible to those used in the MATCH analysis.  We adopted a distance
modulus of 23.0~mag, and used the extinction values from the
\cite{sfd} map.  We also used two sets of age bins, with $\Delta
\log(t)=0.3$ and 0.4~dex, and one metallicity bin, centered at
[Fe/H]$=-$1.5, based on the metallicity estimates from the MATCH fit.
StarFISH was allowed to fit stars from $-$1 to 3 in $(g$-$r)_0$, and
from 17 to 26 in $g_0$ and 17 to 25 in $r_0$, with a bin size of
0.15~mag in each dimension.  The best-fit StarFISH solutions for the
two age-binning schemes are shown in Figure~\ref{fig:starfish}.

\begin{figure*}
\epsscale{1.0}
\plotone{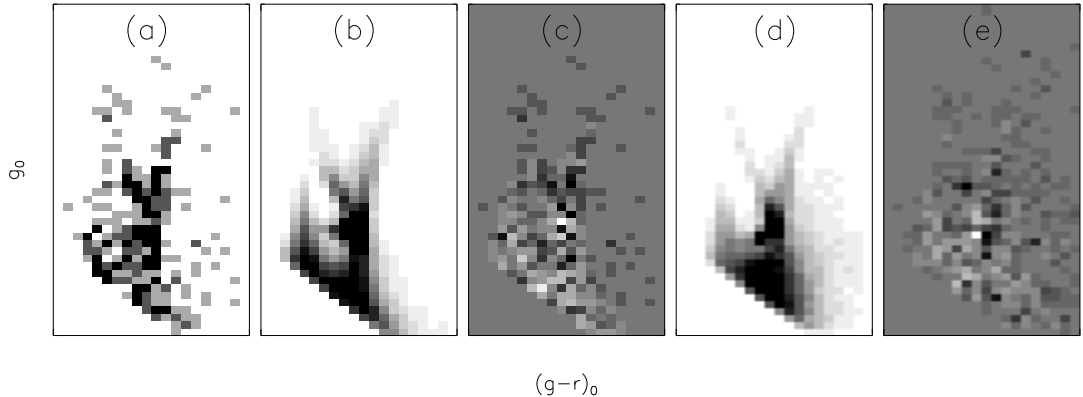}
\caption{ Comparison of data to best MATCH and StarFISH model fits.
{\it (a)} A $(g$-$r)_0$ vs. $g_0$ Hess diagram of the stars within
1\farcm4 of the Leo~T center (i.e., the population used in the SFH
analysis).  {\it (b)} The Hess diagram of the synthetic population
corresponding to the best-fit MATCH solution.  {\it (c)} The residual
Hess diagram after subtracting the best-fit MATCH model from the data.
{\it (d)} Hess diagram of the synthetic population corresponding to
the best-fit StarFISH solution. {\it (e)} The residual Hess diagram
after subtracting the best-fit StarFISH model from the data.  The
sharp cutoff at faint magnitudes is caused by the limiting magnitudes
of $g_0=26$ and $r_0=25$ used with the fits. }
\label{fig:model_hess}
\end{figure*}

Figure \ref{fig:model_hess} shows a comparison of the best-fit models
to the data.  The MATCH and StarFISH results are in statistical
agreement: in both cases, we find extended star formation spanning
ages from a few hundred~Myr to 12~Gyr. The MATCH solution finds the
metallicity to be roughly [Fe/H]$=-$1.5 and indicates two distinct
epochs of star formation, one at 6--12~Gyr and one at 0.4--1~Gyr.
This bimodality is also present in the higher-resolution StarFISH
solution.  These results are also in agreement with the SFH for Leo T
derived by \cite{sdssmatch} based on much shallower SDSS data.
The star formation rates in the two distinct episodes in the higher
resolution fits confirm that the young stellar population represents
approximately 10\% of the stellar mass in Leo T.

The difference between the error bars on the star formation rates in
Figures \ref{fig:matchsfhamr} and \ref{fig:starfish} is due to the
different ways in which they are determined. 
In the case of MATCH, error bars are determined in two ways. First,
for each age bin the minimum and maximum star formation rates are
found in the set of models that qualify as ``good fits''. If all
``good fit'' models give the same star formation rate in a particular
age bin, this error component for that particular bin would be
zero. Second, Monte Carlo simulations are done to assess the
uncertainty associated with the sparseness of sampling of the CMD. For
this, CMDs are drawn from the best-fit CMD model and their SFHs are
determined. The scatter in the star formation rates and metallicities
for each age bin is the second error component. These two error
components are added quadratically to give the final error bars.
StarFISH on the other hand, determines error bars by measuring the
confidence intervals of each amplitude through a systematic
exploration of parameter space in the vicinity of the best fit. 
First, each amplitude is varied in turn while holding all others at
their best-fit values.  This determines the independent uncertainty on
each amplitude.  Next, amplitude pairs that are similar in age or
metallicity are varied together while holding all other amplitudes at
their best-fit values.  This determines the covariant uncertainties
between amplitude pairs that may be partially degenerate.  Finally,
parameter space is explored uniformly by deviating from the best-fit
location along a series of random ``directions'' in parameter space.
We step along each direction, until the fitting statistic increases to
our ``good fit'' limit.  The random-direction search is performed many
thousands of times.  The final error bars are the maximum and minimum
value of each amplitude for all tested parameter space locations whose
fitting statistic fell under the ``good fit'' threshold.  This
difference in approach leads to very different error bars at the young
end, where the StarFISH error bars (Figure \ref{fig:starfish}) are
very large because the upper main-sequence is very poorly populated,
and thus poorly constrained.  On the other hand, this lack of evidence
for very young ($<$100 Myr) stars causes the MATCH error bars in these
bins to be very small (Figure \ref{fig:matchsfhamr}). However, based
on these data some low-level on-going star formation can not be
strictly excluded.

Since the data do not reach the ancient main-sequence turn-off at the
distance of Leo~T, the SFH solutions for ages older than a few Gyr are
based entirely on the morphology of the red giant branch and
horizontal branches.  In particular, at metallicities around
[Fe/H]=$-$1.5, truly ancient core helium-burning stars are predicted to
extend slightly blueward of the canonical ``red clump'' feature (see
the 12.5~Gyr isochrone plotted in Figure~\ref{fig:hess}a).  The
subtlety of this feature in the CMD contributes to the uncertainty of
the derived star formation rate in the oldest age bin.

\section{Summary and conclusions}
\label{sec:conclusions}

Based on deep $g_0$ and $r_0$ photometry obtained with the LBT, we have
studied the stellar populations and structural properties of the Leo T dwarf
galaxy. A comparison of the background-subtracted Hess diagram with
theoretical isochrones confirms the presence of very young stars with ages
between $\sim$200 Myr and 1 Gyr, and an older stellar population ($>5$ Gyr),
the latter with with a likely metallicity of [Fe/H]$\simeq -$1.7.  The stellar
mass in the young population is estimated to be $\sim$10\% of the total
stellar mass.  Based on the apparent magnitude of the RC stars we determine
the distance modulus of Leo T to be 23.1$\pm$0.2, confirming the value from
\cite{LeoT}. The total luminosity of Leo T is estimated to be
$M_{V,total}\simeq -$8.0 mag, almost one magnitude higher than the estimate
from the discovery paper. Using the mass estimate for Leo T from
\cite{simongeha} this implies a mass-to-light ratio of $\sim 60~
M_\sun/L_{V,\sun}$.

A more sophisticated analysis of the photometry using CMD-fitting techniques
yields similar results. The distance determination gives a distance modulus of
$m-M=$23.0$\pm$0.2. CMD fits with two different software packages both
indicate two distinct episodes of star formation, one at $\sim$6 -- 12 Gyr and
one at $\sim$0.4 -- 1 Gyr, although the drop in star formation rate in between
has a low significance. From the CMD fits no strong metallicity evolution is
apparent, with an approximate metallicity of [Fe/H]$\sim -$1.5 at all times.
The metallicity seems to have a small dip around $\log(t)=9.2$, but since the
metallicity constraints at that age come from regions of the CMD where the
stellar models are least reliable, such as the RC and asymptotic giant branch,
this is not significant.

Old and young stars were preferentially selected using CMD-selection boxes in
order to study their individual spatial distributions. The young stars are
more strongly concentrated near the center of Leo T, but both components
display the same spatial extent.  There seems to be a slight offset between
the young and old stars, but from these data this is hardly significant. Since
stars are formed in clusters, young stars are usually spatially more strongly
confined than older stars. This has also been found in other dwarf galaxies
\citep[e.g.][]{hurley99,martinez99,harbeck01,battaglia06,mcconnachie06,martin08}.
The half-light radius for all stars is found to be 120$\pm$7 pc, slightly
smaller than the value from \cite{LeoT}.  

Comparing our results with the properties of the other transition type dwarf
galaxies in the Local group, DDO 210, Phoenix, Pisces, Antlia, Pegasus and Leo
A \citep[e.g.][]{mateo98,grebel01}, shows a lot of similarities. The presence
of both old or intermediate ($>$5 Gyr) age stars and very young ($<$1 Gyr)
stars is common, as is the metallicity of [Fe/H]$\sim-$1.5. A similar apparent
lack of metallicity evolution was found in Leo A by \cite{cole07}. While
in terms of total mass and HI mass Leo T is similar to most other transition
galaxies, it is much fainter. Our luminosity estimate of $M_{V,total}\simeq
-$8.0 is two magnitudes fainter than any of the other transition objects
\citep[e.g.][]{mateo98}. Therefore Leo T remains the least luminous galaxy
with recent star formation known.

We find no evidence for tidal distortion of Leo T. This is not unexpected,
since it is at a large distance from the Milky Way and M31, although based on
the limited information about its orbit it cannot be excluded that Leo T has
come close to the Milky Way some time in the past. However, the lack of
distortion implies that its low mass and luminosity are intrinsic, rather than
the result of interaction or disruption. Thus, the case of Leo T suggests that
isolated halos with masses as low as $10^7 M_\sun$ are able to accrete and/or
retain gas and form stars for at least a Hubble time.

\acknowledgments
The authors thank the LBT Science Demonstration Time (SDT) team for
assembling and executing the SDT program, and the LBC team
and the LBTO staff for their kind assistance. We also thank
Chris Kochanek for help with preparing the manuscript and the anonymous
referee for useful comments and advice.

Facilities: \facility{LBT(LBC)}

\end{document}